\documentclass[conference]{IEEEtran}
\IEEEoverridecommandlockouts

\makeatletter
\def\markboth#1#2{\def\leftmark{\@IEEEcompsoconly{\sffamily}\MakeUppercase{\protect#1}}%
\def\rightmark{\@IEEEcompsoconly{\sffamily}\MakeUppercase{\protect#2}}}
\makeatother

\usepackage[english]{babel}
\usepackage{color}
\usepackage{lipsum}
\usepackage{caption}
\usepackage{cite}
\usepackage[pdftex]{graphicx}
\usepackage{subcaption}
\usepackage{amsmath}
\usepackage{amsfonts}
\usepackage{amssymb}
\usepackage{amsthm}
\usepackage{array}
\usepackage{verbatim}
\usepackage{listings}
\usepackage{algorithm}
\usepackage{algpseudocode}
\usepackage{url}
\usepackage{enumerate}
\usepackage{multirow}

\usepackage{epsfig}
\usepackage{epstopdf}
\usepackage{multicol}
\usepackage[font=footnotesize]{caption}

\renewcommand{\arraystretch}{2}

\newcommand{\bi}{\begin{itemize}}
\newcommand{\ei}{\end{itemize}}
\newcommand{\be}{\begin{equation}}
\newcommand{\ee}{\end{equation}}

\def\beq{\begin{equation}}
\def\eeq{\end{equation}}
\def\beqa{\begin{eqnarray}}
\def\eeqa{\end{eqnarray}}
\def\beqan{\begin{eqnarray*}}
\def\eeqan{\end{eqnarray*}}

\title{\mbox{Resource Sharing in 5G mmWave Cellular Networks}}
\author{{{\bf Mattia Rebato}$^\dagger$, {\bf Marco Mezzavilla}$^\diamond$, {\bf Sundeep Rangan}$^\diamond$, {\bf Michele Zorzi}$^\dagger$ }\\
$^\diamond$NYU Tandon School of Engineering, USA \\
$^\dagger$ University of Padova, Italy \\
emails: $\{$rebatoma, zorzi$\}$@dei.unipd.it, $\{$mezzavilla, srangan$\}$@nyu.edu
}

\begin{document}
\maketitle

\begin{abstract}
In this paper, we discuss resource sharing,
a key dimension in mmWave network design in which
spectrum, access and/or network infrastructure resources can be shared by multiple operators.
It is argued that this sharing paradigm will be essential to fully exploit
the tremendous amounts of bandwidth and the large number of antenna degrees of freedom available in these bands,
and to provide statistical multiplexing to accommodate the highly variable
nature of the traffic. In this paper, we investigate and compare various sharing configurations in order to capture the enhanced potential of mmWave communications. Our results reflect both the technical and the economical aspects of the various sharing paradigms. We deliver a number of key insights, corroborated by detailed simulations, which include an analysis of the effects of the distinctive propagation characteristics of the mmWave channel, along with a rigorous multi-antenna characterization. Key findings of this study include (i) the strong dependence of the comparative results on channel propagation and antenna characteristics, and therefore the need to accurately model them, and (ii) the desirability of a full spectrum and infrastructure sharing configuration, which may result in increased user rate as well as in economical advantages for both service provider. 
\end{abstract}


\section{Introduction}
With the severe spectrum shortage in conventional cellular bands,
millimeter wave (mmWave) frequencies between 10 and 300~GHz have been
attracting growing attention as a possible candidate for next-generation
micro- and pico-cellular wireless networks. The mmWave bands offer orders of magnitude greater spectrum than current cellular allocations and enable very high-dimensional
antenna arrays for further gains via beamforming and spatial multiplexing.
However, due to the unique nature of propagation in these bands,
cellular systems will need to be significantly redesigned~\cite{michele}. Resource sharing is among the most promising approaches to better leverage the potential of mmWave-based frequencies in cellular communications.

Resource sharing has common challenges with heterogeneous networks. Although densification has observable limits for microwave frequencies, it is shown in \cite{Baldemair:15} and \cite{Ghosh:13} that denser deployments are advantageous for mmWave bands because of their different propagation characteristics for non-line of sight (NLoS) and line of sight (LoS) environments\cite{mustafa}.

In the recent literature, we can find contributions that relate to \emph{spectrum} and \emph{infrastructure} sharing in both the \emph{microwave} and the \emph{mmWave} bands, as described in the following.

\textbf{Spectrum sharing:} In the \emph{microwave bands}, where interference is the main limiting factor, competitive and greedy sharing methods might result in severe underutilization of the spectrum, as shown in \cite{Anchora:11}. A viable sensing approach for dynamic inter-operator spectrum sharing for an LTE-A system with carrier aggregation is proposed in \cite{Gerstacker:14}. Under the assumption of partial interference suppression, the optimality of full spectrum sharing is validated by means of simulations in \cite{Zorzi:13}. In the \emph{mmWave bands}, interference avoidance has been shown to give optimum results for WiGig under dense deployments \cite{Zeng:14},~\cite{Liu:13}. However, the directional transmissions typically used at these frequencies allow considerable throughput gains even with blind reuse of frequency bands, e.g., as shown in \cite{Shi:14}, where a ray-tracing model is used to characterize the channel. The authors in \cite{Schulz:14} propose an interference sensing beamforming mechanism in wireless personal access networks (WPANs) that outperforms blind selection algorithms by $15\%-31\%$. In \cite{texas}, based on a simplified channel and antenna model and on a stochastic geometry approach, the authors show that sharing spectrum licenses increases the per-user rate when antennas have narrow beams, and that if network operators share their licenses, they can achieve the same per-user median rate as if each had an exclusive license with more bandwidth.

\textbf{Infrastructure sharing:} Load-aware strategies for the \emph{microwave bands} are proposed in \cite{ElSawy:13}, through an approach based on cognitive spectrum sensing capabilities. In \cite{infra-sha}, the authors investigate the current technological, regulatory, and business landscape from the perspective of sharing network resources, and propose several different approaches and technical solutions for network sharing.
In \cite{texas}, co-location of base stations in the \emph{mmWave bands} is considered. Through a simplified analytical approach, the authors document the potential benefits of infrastructure sharing, although the sensitivity of such results to more accurate channel and antenna models is not discussed. 

In this paper, our goal is to accurately assess the potential of resource sharing in mmWave cellular networks and its sensitivity to mmWave propagation and antenna characteristics. Through a set of simulations that span diverse sharing configurations as well as different channel and antenna models, we deliver some preliminary insights about the feasibility of sharing, as well as its benefits in terms of increased user SINR and rate, as well as of economic advantages. 

The rest of the paper is organized as follows. In Section~\ref{mmWspectrum}, we provide a detailed description of the multi-antenna channel characterization used to obtain our sharing results. In Sections~\ref{configs} and \ref{channels}, we introduce the sharing scenarios investigated and the considered channel and antenna models, respectively. Numerical results are provided in Section~\ref{sims}, and conclusions are drawn in Section~\ref{conclu}. 

\section{MmWave spectrum characterization}
\label{mmWspectrum}
MmWave propagation is characterized by three main challenges, namely (i) \emph{directionality}, obtained from the combination of multiple antennas to increase the intrinsically low transmission range, (ii) \emph{severe shadowing}, because mmWave signals are extremely sensitive to objects, including foliage and the human body, and (iii) \emph{intermittency}, where obstacles can lead to much more dramatic fluctuations of the channel gain, resulting in frequent and sudden drops (e.g., people passing through the area of coverage)~\cite{fol1, fol2, ted_book}.

Since these effects play a key role in determining the performance in a mmWave scenario, it is particularly important to carefully characterize the propagation phenomena, and the directional gains introduced by multi-antenna schemes. 

We start with the traditional definition of link budget
\be
P_{\text{RX}} = P_{\text{TX}} + G_{\text{BF}} - PL - \xi,
\ee
where $P_{\text{RX}}$ and $P_{\text{TX}}$ are the received and transmitted power in dBm, respectively; $G_{\text{BF}}$, $PL$ and $\xi$ are the beamforming gain, pathloss and shadowing in dB, respectively~\cite{ted_book}.

\subsection{MmWave propagation}
From the measurement campaign carried out in a real dense urban environment and reported in~\cite{mea1,mea2,mea3}, pathloss can be modeled with three states: LoS, NLoS and outage. Each link is characterized by the channel state probabilities $p_{\mathrm{LoS}}$, $p_{\mathrm{NLoS}}$ and $p_{\mathrm{out}}$, which are expressed in terms of the distance $d$ between the user equipment (UE) and the base station (BS) as follows:
\be
\begin{matrix}
\vspace{-0.3cm}
p_{\mathrm{out}}(d)=\mathrm{max}(0,1-e^{-a_{\mathrm{out}}d+b_{\mathrm{out}}})\\ \vspace{-0.3cm}
p_{\mathrm{LoS}}(d)= (1-p_{\mathrm{out}}(d))e^{-a_{\mathrm{LoS}}d}
\\  
p_{\mathrm{NLoS}}(d)= 1-p_{\mathrm{out}}(d)-p_{\mathrm{LoS}}(d),
\end{matrix}
\ee
where $a_{\mathrm{out}}=0.0334$ m$^{-1}$, $b_{\mathrm{out}}=5.2$ and $a_{\mathrm{LoS}}=0.0149$ m$^{-1}$ (all these values are taken from~\cite{mustafa} assuming a carrier frequency of 28 GHz). 
On the other hand, the pathloss is given by:
\be
PL(d)[dB] = \alpha + \beta 10 \log_{10}(d),
\ee
and the log-normal shadowing is $\xi\sim N(0,\sigma^2)$, where ($\alpha,\beta,\sigma$) $=$ ($61.4$, $2$, $5.8$ dB) for LoS and ($72$, $2.92$, $8.7$ dB) for NLoS~\cite{mustafa}. 

\subsection{MIMO channel}
The channel is assumed to be composed of a random number $K$ of clusters, each corresponding to a macro-level scattering path. At the receiver, the number of estimated clusters is given as the maximum value between 1 and a Poisson random variable whose mean $\lambda$ is related to the carrier frequency as explained in~\cite{mustafa}. For each cluster $k$, the number of sub-paths is modeled as an integer random variable $L_k$ uniformly distributed in $\{1,\dots,10\}$.
Given a set of clusters and of sub-paths for a channel, we can compute the channel matrix as:
\begin{equation}
\textbf{H}(t,f)=\sum_{k=1}^{K}\sum_{l=1}^{L_k}g_{kl}(t,f) \textbf{u}_{Rx}(\theta^{Rx}_{kl},\phi^{Rx}_{kl}) \textbf{u}^*_{Tx}(\theta^{Tx}_{kl},\phi^{Tx}_{kl}),
\label{channel_matrix}
\end{equation}
where:
\begin{itemize}
\item {$L_k$}: number of sub-paths in cluster $k$;
\item {$g_{kl}(t,f)$}: small-scale fading over time and frequency;
\item {$ \textbf{u}_{Rx}(\cdot)$}: spatial signature vector of the receiver;
\item {$ \textbf{u}_{Tx}(\cdot)$}: spatial signature vector of the transmitter.
\end{itemize}
Spatial signatures are computed with horizontal and vertical angles of arrival (AoA) $\theta^{Rx}_{kl},\phi^{Rx}_{kl}$, and horizontal and vertical angles of departure (AoD) $\theta^{Tx}_{kl},\phi^{Tx}_{kl}$, where $k = 1, \dots, K$ is the cluster index and $l = 1, \dots, L_k$ is the sub-path index within the cluster.

The small-scale fading is generated based on the number of clusters, the number of sub-paths in each cluster, the Doppler shift, the power spread, and the delay spread, as:
\begin{equation}
g_{kl}(t,f)=\sqrt{P_{lk}}e^{j 2\pi f_{d}cos(\omega_{kl} )t}e^{-j2\pi \tau _{kl}f},
\label{scale_fading}
\end{equation}

where:

\begin{itemize}
\item {$P_{kl}$}: power spread of sub-path $l$ in cluster $k$;
\item {$f_{d}$}: maximum Doppler shift;
\item {$\omega_{kl}$}: angle of arrival of sub-path $l$ in cluster $k$ with respect to the direction of motion;
\item {$\tau _{kl}$}: delay spread of sub-path $l$ in cluster $k$;
\item {$f$}: carrier frequency.
\end{itemize}

The power spread $P_{kl}$ is obtained by following~\cite{mathew}:
\be
P_{kl} = \frac{P_{kl}^\prime}{\sum P_{kl}^\prime},
\ee
\be
P_{k}^\prime = \frac{U_k^{r_{\tau}-1}10^{-0.1 Z_k+V_k}}{L_k},
\ee
where $U_k \sim U[0,1]$, $V_k \sim U[0,0.6]$ and $Z_k \sim N(0,\zeta^2)$, while parameters $r_\tau$  and $\zeta$ are found in~\cite{mustafa}.

\subsection{Beamforming}
Due to high pathloss, multiple-input and multiple-output (MIMO) systems with beamforming techniques are essential to ensure an acceptable range of communication in mmWave networks.
Using the channel matrix $\textbf{H}$ computed in Eq.~(\ref{channel_matrix}), the beamforming gain from transmitter $i$ to receiver $j$ is given as:
\be
G_{ij} = |\textbf{w}^*_{Rx_{ij}} \textbf{H}_{ij} \textbf{w}_{Tx_{ij}} |^2
\label{bf_comp}
\ee
where $\textbf{w}_{Tx_{ij}} \in \mathbb{C}^{n_{Tx}}$ is the beamforming vector of transmitter $i$ when transmitting to receiver $j$, and $\textbf{w}_{Rx_{ij}} \in \mathbb{C}^{n_{Rx}}$ is the beamforming vector of receiver $j$ when receiving from transmitter $i$. 
Both vectors are complex, with length equal to the number of antenna elements in the array.
These beamforming vectors have been computed following the procedure in~\cite{antenna_book}.
In each simulation, we assume to be able to steer towards any angle, so that the transmitter and receiver beamforming patterns are always perfectly aligned. 

\begin{figure}[t!]
\centering
    \includegraphics[width=0.75\columnwidth]{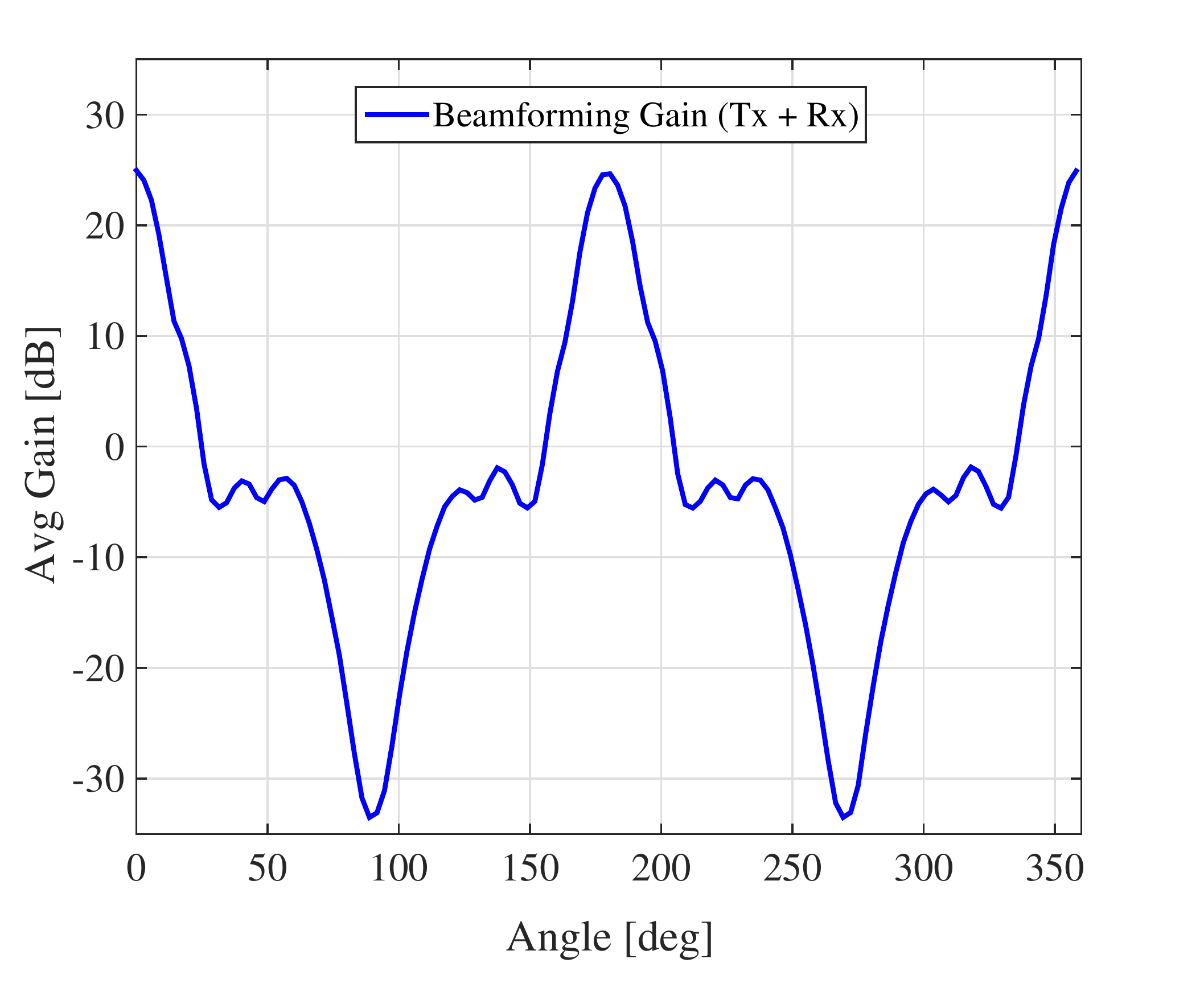}
      \caption{Aggregate transmit plus receive beamforming gain.}
      \label{bf_gain}
\end{figure}
As an example, we report the aggregate beamforming gain in Fig.~\ref{bf_gain}, which includes both transmit and receive sides and refers to a $64\times16$ MIMO system.
We note that no blocking reflectors are assumed here, which are usually placed to avoid the symmetrical lobe on the backside of the antenna array. In Section \ref{sims}, we will show results for both options. 

Thanks to our detailed channel and antenna characterization, we can compute the signal-to-interference-plus-noise-ratio (SINR) between transmitter $i$ and receiver $j$ as:
\be
SINR_{ij} = \frac{\frac{P_{Tx,ij}}{PL_{ij}}G_{ij}}{\sum_{k \neq i} \frac{P_{Tx,kj}}{PL_{kj}}G_{kj} + BW \times N_0}
\label{equation_sinr}
\ee
where $k$ represents each interfering link, $BW$ is the total bandwidth, and $N_0$ is the thermal noise.

The expression in (\ref{equation_sinr}) is useful to evaluate the system performance in terms of \emph{SINR coverage}, which is defined as the probability that the SINR experienced by a user is above a certain threshold $T$, as follows:   

\be
C_{SINR} (T) = \mathbb{P}[SINR > T].
\ee

Another important metric that we consider in our performance comparisons is the \emph{rate coverage}, which represents the distribution of the data rate achieved by each user, and is formally
defined as the probability that the rate of a user is greater than a threshold $\rho$:
\be
C_{rate}(\rho) = \mathbb{P}\left[\frac{\alpha BW}{N} \log_2(1 + SINR) > \rho\right]
\ee
where $N$ is the number of users connected to a base station and $\alpha=0.5$ is the half duplex factor.

\section{Sharing Configurations}
\label{configs}

\begin{figure*}[t!]
\centering
    \includegraphics[width=0.8\textwidth]{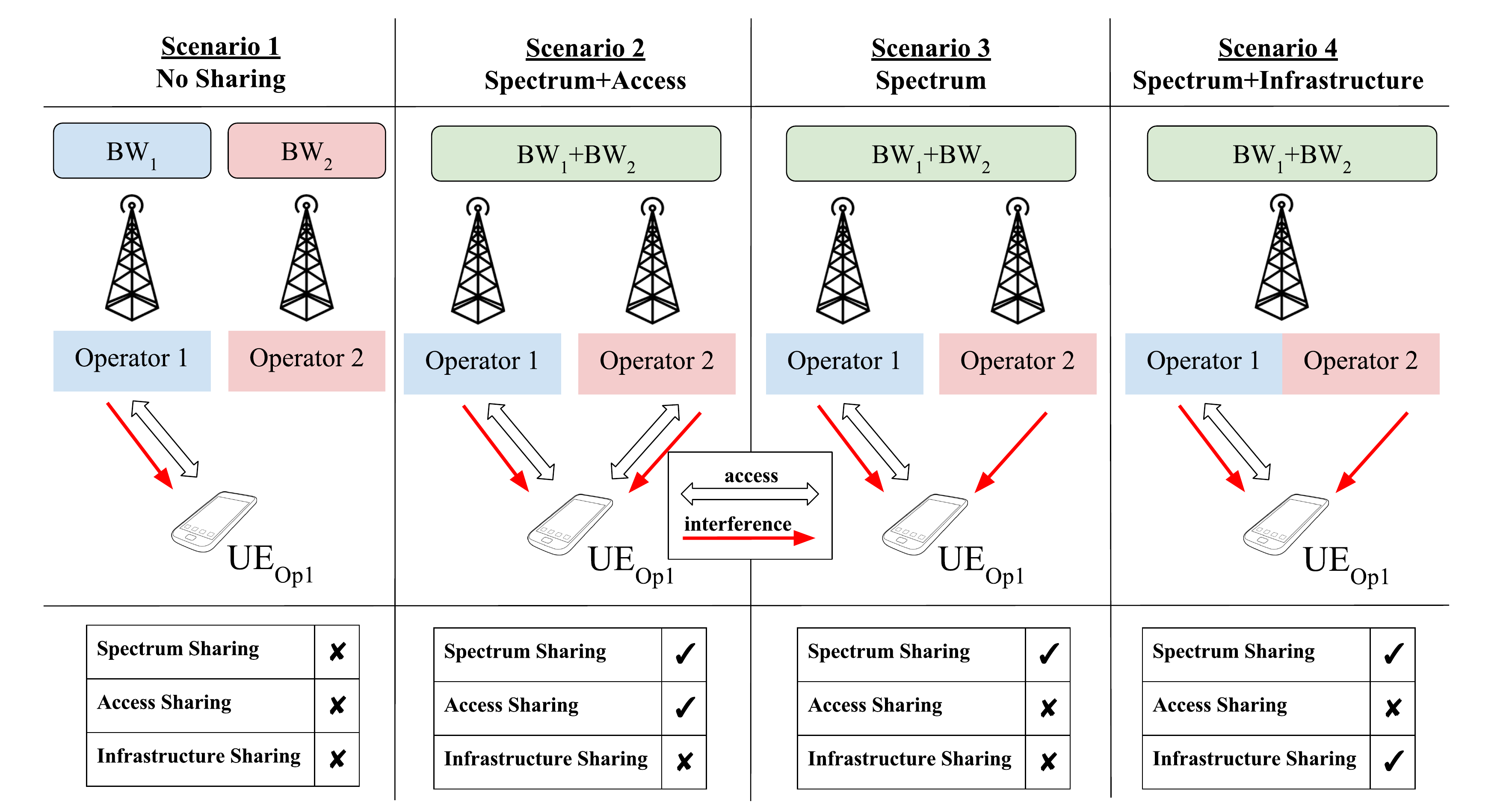}
      \caption{Considered \{spectrum, access, infrastructure\} sharing configurations. }
      \label{shar_conf}
\end{figure*}

As in~\cite{texas}, in this paper we want to compare different sharing configurations, and to derive some insights about the feasibility and performance benefits of resource sharing in mmWave networks. The scenarios considered in this paper are reported in Fig.~\ref{shar_conf}, and described in the following.\footnote{In this paper, we implicitly refer to the common case of two operators. Extension to the case of more than two operators is straightforward.}

\textbf{Scenario 1 (No Sharing):} This is our \emph{benchmark}, where operators transmit in their own bands, which are orthogonal to each other, and utilize their own network infrastructures.
Moreover, users can only connect to the operator they subscribe to.
This is a traditional network architecture, used to assess the performance improvements obtained by the various sharing options.

\textbf{Scenario 2 (Spectrum+Access):} Network providers share the same spectrum, and thus have a wider available bandwidth.
They operate on separate infrastructures, but users can be associated to any operator.
Due to the intrinsic complexity of this scenario, where full coordination is needed among the service providers to enable access sharing, it serves as an upper bound for the more realistic scenarios illustrated below. 
From a mathematical point of view, this case corresponds to the previous one with twice the densities of UEs and BSs and a double amount of bandwidth.

\textbf{Scenario 3 (Spectrum):} Network providers share the same spectrum but operate on separate infrastructures. Unlike in the previous scenario, users can only connect to their associated operator. 

\textbf{Scenario 4 (Spectrum+Infrastructure):} This is similar to Scenario 3, but with co-located BSs, so that, besides sharing their spectrum, the operators also use a common network infrastructure.
More precisely, each BS site hosts one antenna for each operator. Note that this can be achieved if each of the two operators acquires half as many BS sites and shares them with the other, thereby obtaining a dense infrastructure (with improved SINR and rate coverage) at a reduced cost compared to the case of separate infrastructures. 

\section{Channel and Antenna Models}
\label{channels}

In order to assess to what extent the benefits of spectrum sharing may depend on the specific propagation and antenna characteristics, we consider the following four models.

\textbf{Model 1: } This model follows the characterization provided in~\cite{texas}, which presents a \emph{rect}-shape model of the beamforming gain with maximum and minimum gains $\mathrm{G_{max}}$ and $\mathrm{G_{min}}$, respectively, and with half beamwidth $\theta_b$. To make it comparable with the other considered models, we select these parameters based on Fig.~\ref{bf_gain}, as $\theta_b=28^{\circ}, \mathrm{G_{max}}=26$ dB and $\mathrm{G_{min}}=-4$ dB. 
The pathloss is modeled through a constant times the inverse power of the distance, with parameters  related to the LoS and NLoS conditions as described in~\cite{texas}.

\textbf{Model 2: } This channel model is fully characterized based on the measurements reported in~\cite{mustafa},~\cite{mathew}, as detailed in Section~\ref{mmWspectrum}. In addition, we provide a precise multi-antenna gain characterization, which relies on beamforming vectors computed following~\cite{antenna_book}, as illustrated in Eq.~(\ref{bf_comp}).

\textbf{Model 3: } This model coincides with the previous one, with the only exception that the back lobe is removed, thus resulting in less interference and better overall performance.
As previously described, this can be achieved with a metal plane on the backside of the antenna array.

\textbf{Model 4: } Finally, we propose a \emph{mixed} model, where the pathloss is based on Model 1, whereas the multi-antenna is fully characterized by matrix $\textbf{H}$ and beamforming vectors $\textbf{w}_{Tx},\textbf{w}_{Rx}$ as in Model 2.
This model makes it possible to evaluate the effect of the simplified antenna characterization of~\cite{texas}, and to assess the sensitivity of the sharing results to the antenna model used.

\section{Simulation Results}
\label{sims}

All our simulation parameters are reported in Table~\ref{table_of_parameters}, where we assume that each of the two networks owns a license for 500 MHz.
In Scenarios 2, 3 and 4, both networks share their spectrum licenses, so that the same 1 GHz of total bandwidth is available to both operators.

Our simulations follow a Monte Carlo approach, in which many independent experiments are repeated to empirically derive statistical quantities of interest. Each experiment consists in (i) deploying UEs and BSs according to two Poisson point processes, as done in~\cite{texas} and~\cite{ElSawy:13}, (ii) establishing UE-BS associations according to a minimum pathloss criterion, and (iii) computing the SINR of the user at the center of the area. The SINR statistics is estimated based on 10$^4$ repetitions of this procedure.

\begin{table}[h!]
\centering
 \renewcommand\arraystretch{1.3}
\renewcommand\tabcolsep{1.5pt}
\begin{tabular}{|c|c|c|}
\hline
\textbf{Notation} & \textbf{Value} & \textbf{Description} \\
\hline
\hline
$M$ & 2 & Number of operators\\
\hline
$\lambda_{UE}$& 200  & UE density per $km^2$ \\
\hline
$\lambda_{BS}$& 30 & BS density per $km^2$ \\
\hline
$A $ & 1 km$^2$ & Area of the simulations \\ 
\hline
$P_{Tx} $ & 30 dBm & Transmitting power \\
\hline
$f$ & 28 GHz & Carrier frequency \\
\hline
$BW$ & 1 GHz & Total bandwidth\\
\hline
$NF$ & 7 dB & Noise figure\\
\hline
$n_{Tx}$ & 64 elements &$8\times 8$ UPA TX antennas \\  
 \hline
$n_{Rx}$ & 16 elements &$4\times 4$ UPA RX antennas \\  
\hline
\end{tabular}
\caption{Simulation parameters}
\label{table_of_parameters}
\end{table}

\begin{figure}[t!]
\centering
    \includegraphics[width=0.76\columnwidth]{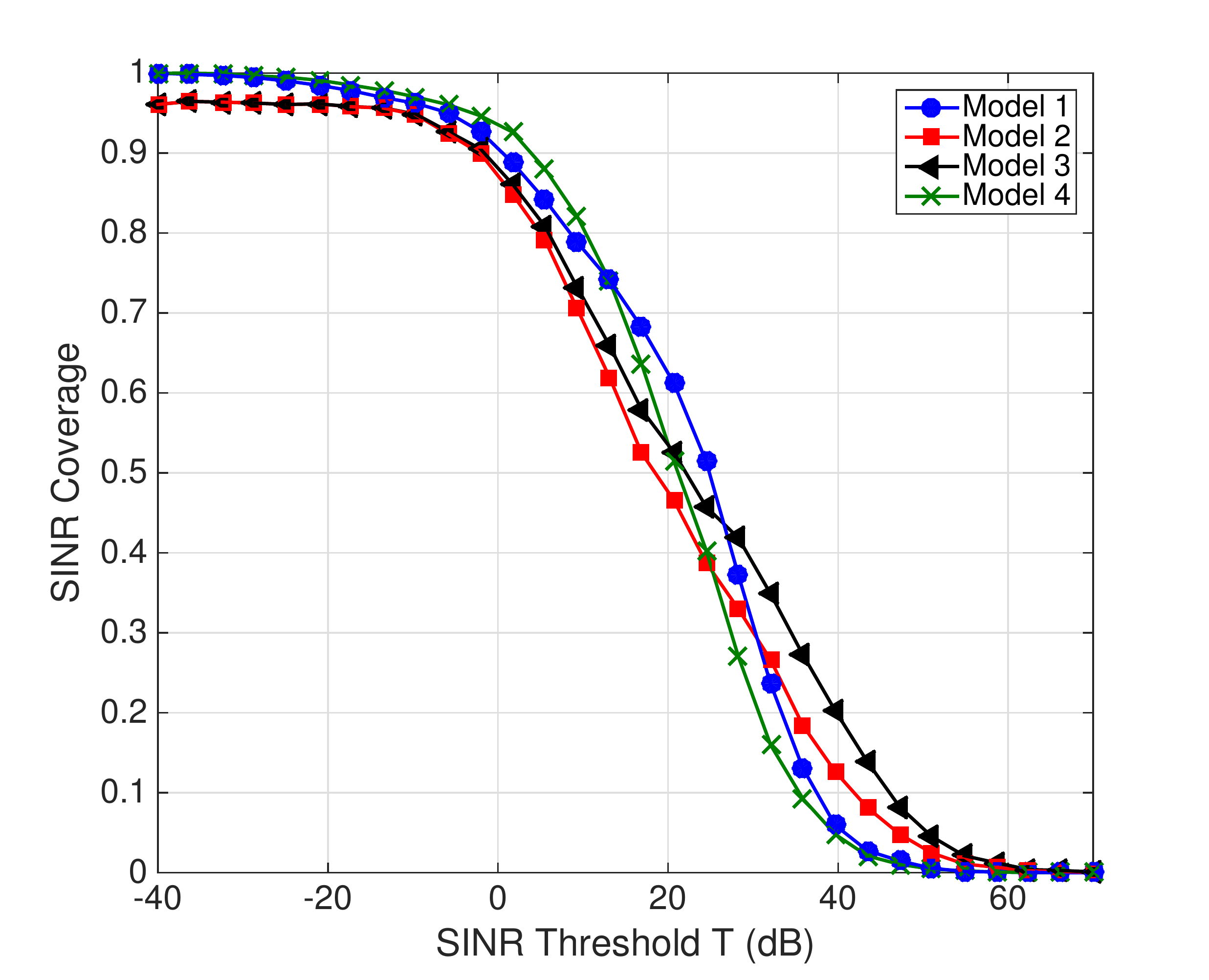}
      \caption{SINR coverage for Scenario 1, low node density ($\lambda_{BS}=30$, $\lambda_{UE}=200$), all models.}
      \label{sce1}
\end{figure}

\begin{figure}[t!]
\centering
    \includegraphics[width=0.76\columnwidth]{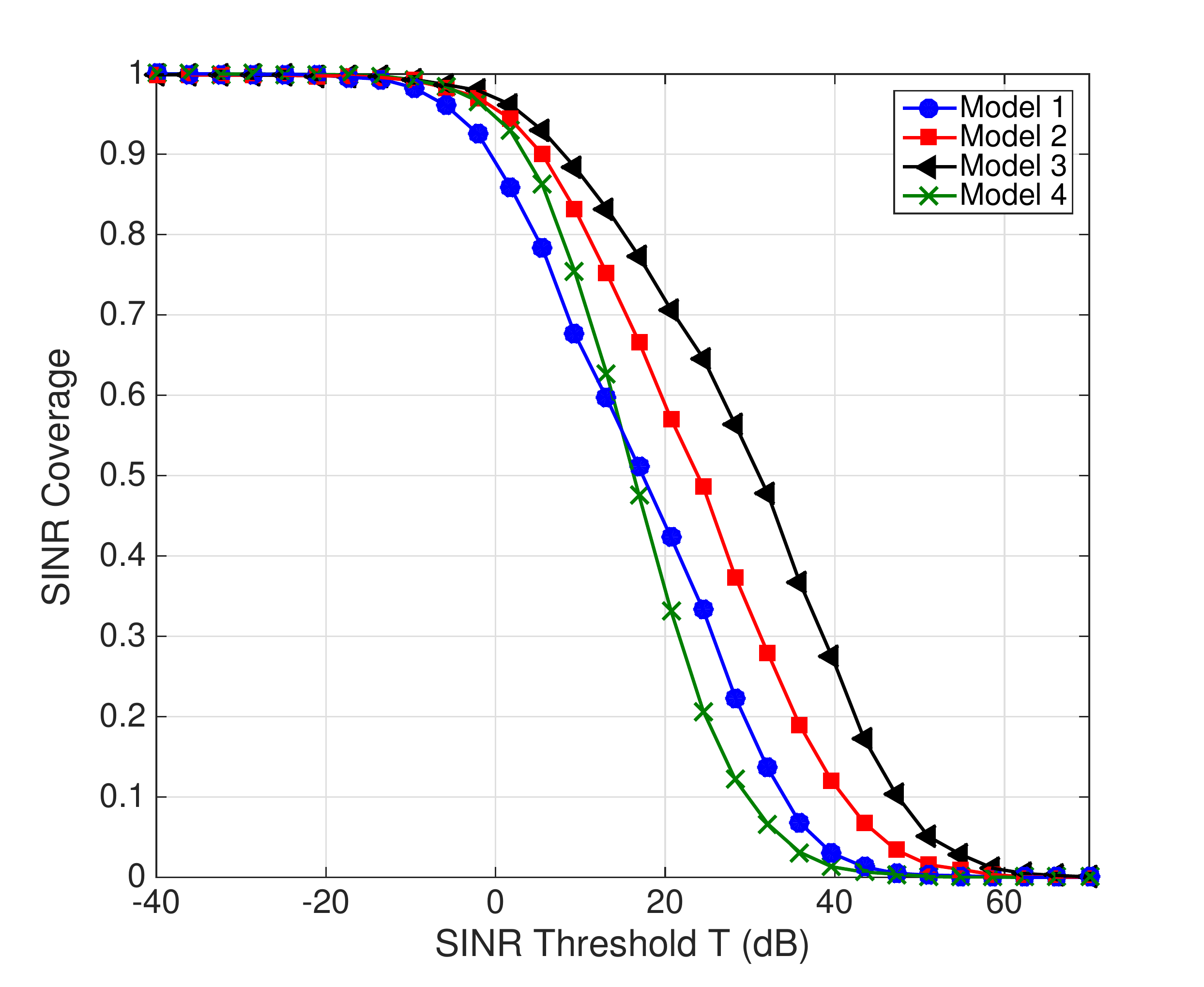}
      \caption{SINR coverage for Scenario 2, high node density ($\lambda_{BS}=60$, $\lambda_{UE}=400$), all models.}
      \label{sce2}
\end{figure}

First of all, we compare in Figs.~\ref{sce1} and \ref{sce2} the performance results obtained in Scenario 1 and Scenario 2 with the four different channel models described in Section~\ref{channels}. This comparison highlights the ultimate potential of sharing, as Fig.~\ref{sce1} shows the benchmark performance already achievable in current systems, whereas Fig.~\ref{sce2} reports the upper bound that could be achieved if full sharing were available. 

Based on these results, we highlight the following interesting points:

\textbf{1)} The sensitivity of the SINR coverage results to the channel and antenna model, which is rather modest in the no-sharing case of Fig.~\ref{sce1}, is significantly increased when sharing is allowed. This is shown by the dispersion of the curves in the two figures, which in the sharing case may be as high as almost 20 dB. This clearly highlights that, while simplified channel and antenna models can be adequate to study traditional systems, they may become less so for the study of spectrum sharing scenarios, where using an approximate channel may lead to severe underestimation of the achievable performance.

\textbf{2)} Model 3 always outperforms Model 2, due to the fact that, by removing the back lobe in Model 3, we are reducing the interference, thus increasing the SINR coverage.
This back-lobe suppression can be easily implemented if multiple uniform planar array (UPA) antennas are applied at the transmit side (e.g., three UPA antennas with orientations shifted by 120 degrees each).


\textbf{3)} Interference-related effects also emerge when comparing the trends in Scenario 1 and Scenario 2. Fig.~\ref{sce2} depicts a higher density topology\footnote{Note that, for the full sharing scenario, the densities are doubled with respect to those given in Table~\ref{table_of_parameters}.}, which results in a higher overall level of interference (more transmitters) but also in a better quality of the intended signal (the serving BS is closer). The combined effect of these two trends leads to non-obvious behaviors, where for example we observe that (i) the dispersion of the SINR values is wider for Models 2 and 3 based on~\cite{mustafa} than for Models 1 and 4 based on~\cite{texas}, and (ii) sharing leads to an SINR coverage enhancement for Models 2 and 3, and to its degradation for Models 1 and 4. This shows that the pathloss models have a significant impact on the behavior of the curves, which depends on the specifics of the models and needs to be examined in more detail. We remark that a degradation in SINR coverage may still corresponds to significantly better rate coverage, when the reduction in spectral efficiency is overweighed by the increase in available bandwidth (which is doubled).

\textbf{4)} The only difference of Model 4 with respect to Model 1, as mentioned before, is a detailed multi-antenna gain characterization, which results in less beamforming gains due to the more realistic, \emph{lobe}-shaped, radiation pattern shown in Fig.~\ref{bf_gain}, as opposed to the \emph{rect}-shaped approximation presented in~\cite{texas}, which only takes two values, $\mathrm{G_{max}}$ and $\mathrm{G_{min}}$. According to the results of Figs.~\ref{sce1} and \ref{sce2}, this leads to a better cell-edge coverage, although a significant fraction of the users will experience a decreased SINR.
 
\textbf{5)} Models 2 and 3 show less coverage than Models 1 and 4 for lower SINR thresholds and no sharing. This is due to the fact that an outage probability, $p_{\mathrm{out}}$, is included in the blockage model, which clearly shows in a low-density scenario (see the left part of the curves in Fig.~\ref{sce1}), but vanishes for higher densities (see Fig.~\ref{sce2}).

\begin{figure}[t!]
\centering
    \includegraphics[width=0.76\columnwidth]{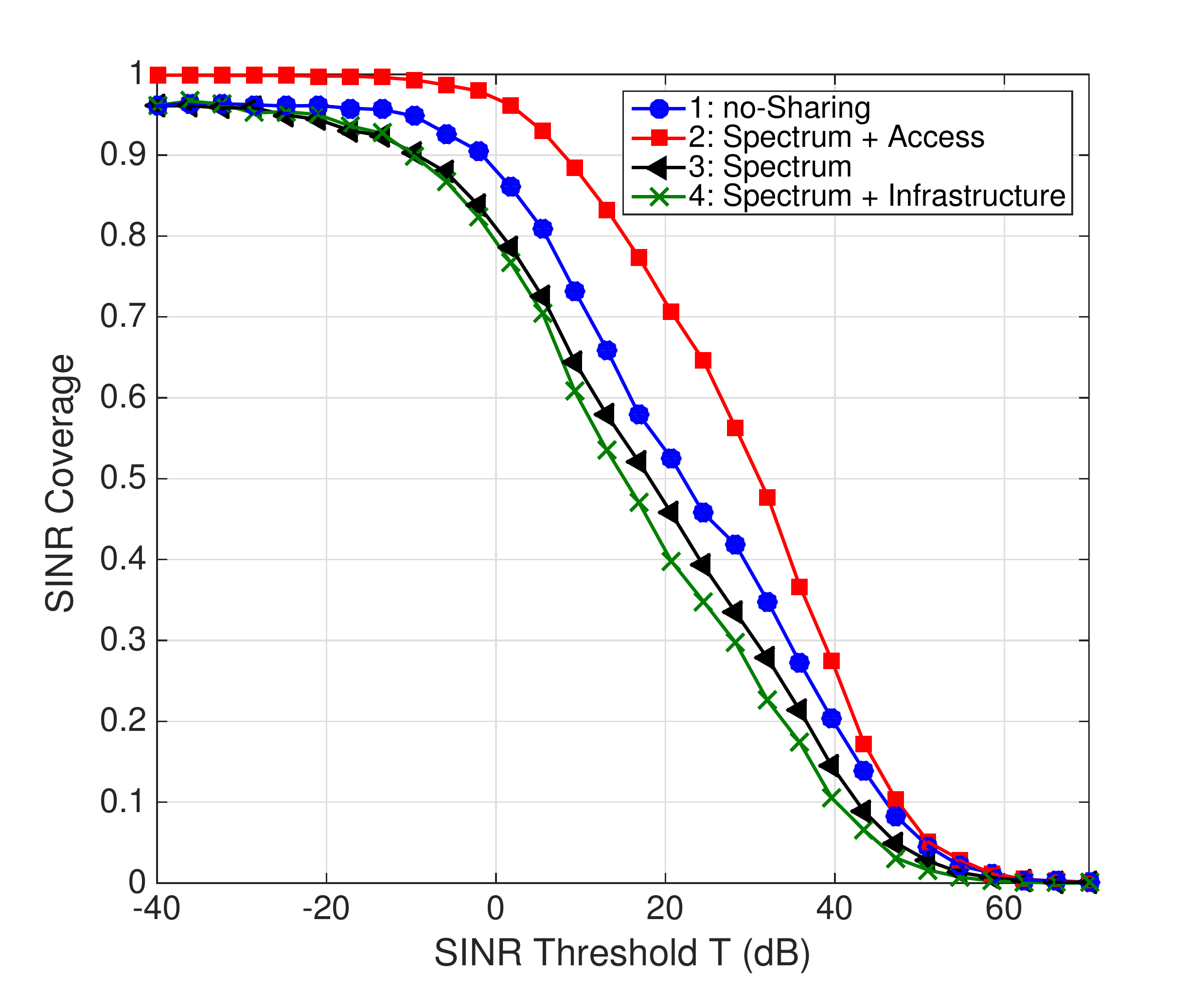}
      \caption{SINR coverage for Model 3, all scenarios.}
      \label{mods_pc}
\end{figure}
\begin{figure}[t!]
\centering
    \includegraphics[width=0.76\columnwidth]{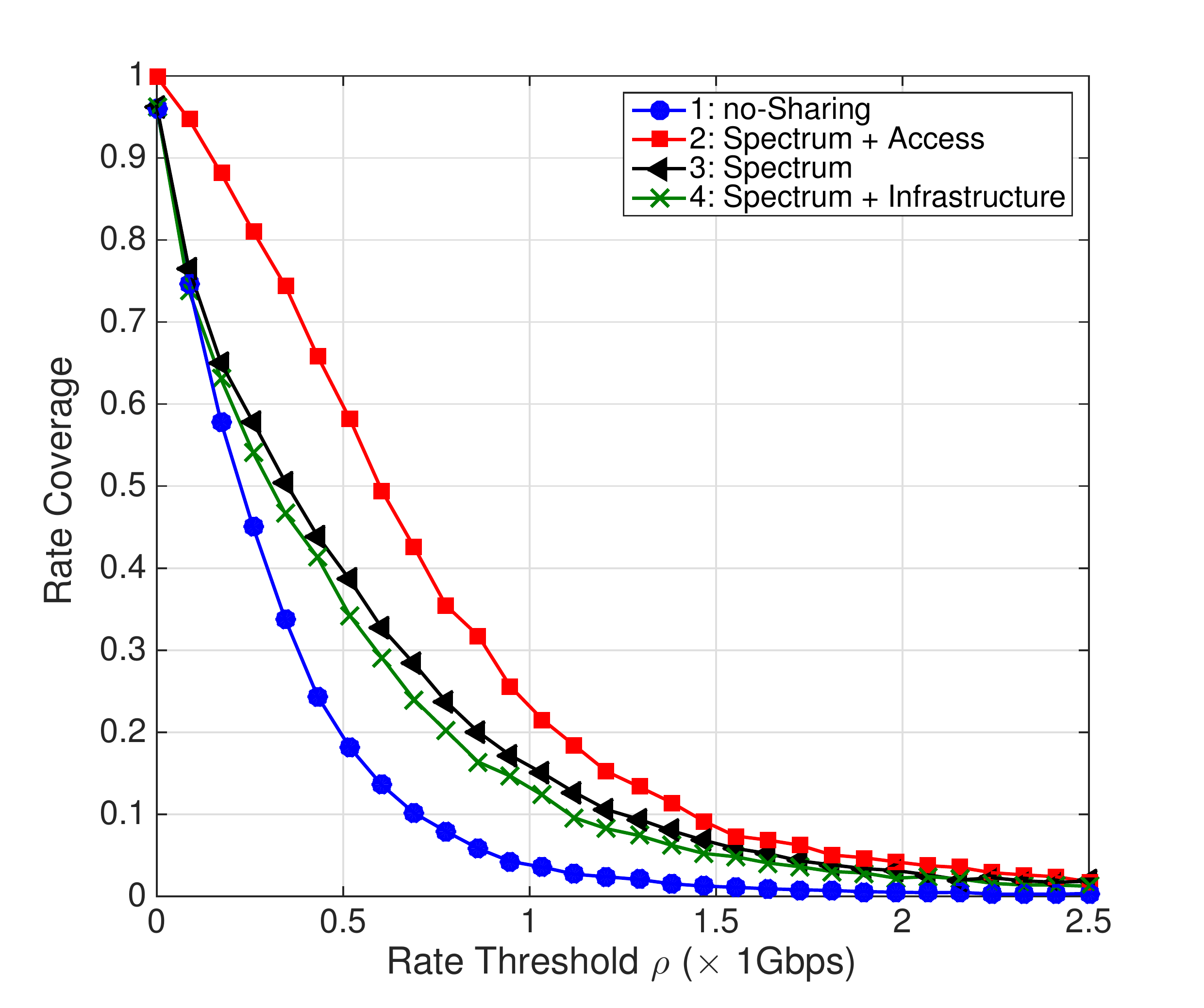}
      \caption{Rate coverage for Model 3, all scenarios.}
      \label{mods_rate}
\end{figure}

For a more detailed comparison of the performance of the various sharing options, we plot all scenarios in Figs.~\ref{mods_pc} and~\ref{mods_rate}, focusing on the most realistic channel and antenna representation, i.e., Model 3. 

Some additional insights can be summarized as follows: 

\textbf{6)} In terms of SINR coverage, a full spectrum and access sharing configuration, i.e., Scenario 2, outperforms any other configuration. This, as explained in the previous section, serves as an upper bound because of the access sharing complexity, and is due to the fact that each user has more association opportunities and therefore better signal quality. 

\textbf{7)} Configurations with spectrum sharing and closed access, like Scenarios 3 and 4, show less SINR coverage than the no-sharing case of Scenario 1 (see Fig.~\ref{mods_pc}), because of the increased interference, but outperform it in terms of rate coverage (see Fig.~\ref{mods_rate}), because of the increased available bandwidth. 

\textbf{8)} Finally, we can note that Scenario 4, which is the most favorable setting for the operators from an economic perspective\footnote{Network providers share the infrastructure costs.}, presents only a slight degradation compared to Scenario 3. While this degradation was expected because of the constraints on BS placement imposed by infrastructure sharing, it is interesting to observe that the performance loss is actually quite limited, especially in terms of rate. 
This shows that infrastructure sharing has the potential to provide an economical means to densify the network, achieving a performance level similar to what could be obtained with separate infrastructures, while using only half as many BS sites. Based on the results shown in this section, we can draw the following key conclusions:

\textbf{(i)} Operators that share their licenses (frequency bands) can access more resources, thus providing higher rate for the average user of both providers. 

\textbf{(ii)} Full spectrum and access sharing outperforms any other sharing scheme, in terms of both SINR and rate coverage, as users have more opportunities to find a BS in-range (because of the increased BS density) and can achieve higher data rates (because of the increased bandwidth).

\textbf{(iii)} Scenarios with co-located antennas, namely infrastructure sharing, can obtain the performance gains achievable by network densification while incurring a significantly reduced deployment cost.

\textbf{(iv)} While a simplified channel and antenna characterization (e.g., as used in~\cite{texas}) can be used to qualitatively assess the performance trends of sharing techniques, accurate models as those used in this paper are critical in order to be able to precisely quantify the achievable gains.

\section{Conclusions}
\label{conclu}
In this paper, we showed how resource sharing represents a solution to better leverage the potential of mmWave technology for cellular networks, where very large bandwidths and many antenna degrees of freedom are available. Through some detailed simulation results based on accurate channel and antenna models, we have characterized the benefits of resource sharing in these bands.
Some key findings of our simulation study include (i) the need to accurately model channel propagation and antenna characterization to better capture the resource sharing capabilities, and (ii) the desirability of a full spectrum and infrastructure sharing configuration, which results in increased user rate as well as in economical advantages for both service providers.

\bibliographystyle{IEEEtran}
\bibliography{biblio.bib}

\end{document}